\documentclass[prm,twocolumn,floatfix,superscriptaddress,amsmath,citeautoscript,aps,longbibliography]{revtex4-1}
\usepackage{blindtext}
\usepackage{graphicx}% Include figure files
\usepackage{tabularx}
\usepackage{array}
\usepackage{booktabs}
\usepackage{dcolumn}% Align table columns on decimal point
\usepackage{bm}% bold math
\usepackage{color}
\usepackage[normalem]{ulem}% for \sout
\usepackage{natbib}
\newcommand{\stkout}[1]{\ifmmode\text{\sout{\ensuremath{#1}}}\else\sout{#1}\fi}
\definecolor{magenta}{cmyk}{ 0, 1, 0,0}
\usepackage{amsmath}
\usepackage{float}
\usepackage{bm}
\usepackage{wasysym}
\usepackage{empheq}

\usepackage{hyperref}  % for cite and reference
\hypersetup{hypertex=true,
            colorlinks=true,
            linkcolor=blue,
            anchorcolor=blue,
            citecolor=blue}

\newcommand{\DNG}{DyNi$_5$Ge$_3$}

\begin{document}

\title{Successive magnetic orderings in the Ising spin chain magnet DyNi$_5$Ge$_3$ }

\date{\today}
\author{H. Ge}
\thanks{These authors have contributed equally to this work.}
\affiliation{Department of Physics, Southern University of Science and Technology, Shenzhen 518055, China}

\author{L. Zhang}
\thanks{These authors have contributed equally to this work.}
%\affiliation{Department of Physics, Southern University of Science and Technology, Shenzhen 518055, China}
\affiliation{Ganjiang Innovation Academy, Chinese Academy of Sciences, Ganzhou 341000, China}

\author{N. Zhao}
\affiliation{Department of Physics, Southern University of Science and Technology, Shenzhen 518055, China}

\author{J. Yang}
\affiliation{Department of Chemistry, Southern University of Science and Technology, Shenzhen 518055, China}

\author{L. Wang}
\affiliation{Department of Physics, Southern University of Science and Technology, Shenzhen 518055, China}
\affiliation{Shenzhen Institute for Quantum Science and Engineering,Shenzhen 518055, China}

\author{L. Zhou}
\affiliation{Department of Physics, Southern University of Science and Technology, Shenzhen 518055, China}

\author{Y. Fu}
\affiliation{Department of Physics, Southern University of Science and Technology, Shenzhen 518055, China}

\author{T.~T. Li}
\affiliation{Department of Physics, Southern University of Science and Technology, Shenzhen 518055, China}

\author{Z.~M. Song}
\affiliation{Department of Physics, Southern University of Science and Technology, Shenzhen 518055, China}

\author{F. Ding}
\affiliation{Department of Physics, Southern University of Science and Technology, Shenzhen 518055, China}

\author{J.~B. Xu}
\affiliation{Department of Physics, Southern University of Science and Technology, Shenzhen 518055, China}

\author{Y.~F. Zhang}
\affiliation{Department of Physics, Southern University of Science and Technology, Shenzhen 518055, China}

\author{S.~M. Wang}
\affiliation{Department of Physics, Southern University of Science and Technology, Shenzhen 518055, China}

\author{J.~W. Mei}
\affiliation{Department of Physics, Southern University of Science and Technology, Shenzhen
  518055, China}
\affiliation{Shenzhen Institute for Quantum Science and Engineering,Shenzhen
  518055, China}
\affiliation{Shenzhen Key Laboratory of Advanced Quantum Functional Materials
   and Devices, Southern University of Science and Technology, Shenzhen 518055, China}

\author{X. Tong}
\affiliation{Institute of High Energy Physics, Chinese Academy of Sciences (CAS), Beijing 100049, China}
\affiliation{Spallation Neutron Source Science Center, Dongguan 523803, China}

\author{P. Miao}
\affiliation{Institute of High Energy Physics, Chinese Academy of Sciences (CAS), Beijing 100049, China}
\affiliation{Spallation Neutron Source Science Center, Dongguan 523803, China}

\author{H. He}
\affiliation{School of Chemical Science, University of Chinese Academy of Sciences (UCAS), Beijing 100190, China}

\author{Q. Zhang}
\email{zhangq6@ornl.gov}
\affiliation{Scattering Division, Oak Ridge National Laboratory, Oak Ridge, Tennessee 37831, USA}

\author{L.~S. Wu}
\email{wuls@sustech.edu.cn}
\affiliation{Department of Physics, Southern University of Science and Technology, Shenzhen 518055, China}
\affiliation{Shenzhen Key Laboratory of Advanced Quantum Functional Materials and Devices,
Southern University of Science and Technology, Shenzhen 518055, China}

\author{J.~M. Sheng}
\email{shengjm@sustech.edu.cn}
\affiliation{Academy for Advanced Interdisciplinary Studies, Southern University of Science and Technology,
Shenzhen 518055, China}
\affiliation{Department of Physics, Southern University of Science and Technology,
Shenzhen 518055, China}

\date{\today}

\date{\today}

\begin{abstract}
In this report, we investigated a new rare earth based one-dimensional Ising spin chain magnet~\DNG~by means of magnetization, specific heat and powder neutron diffraction measurements. Due to the crystalline electrical field splitting, the magnetic Dy ions share an Ising like ground doublet state. Owning to the local point symmetry, these Ising moments form into two canted magnetic sublattices, which were further confirmed by the angle-dependent magnetization measurement. In zero fields, two successive antiferromagnetic phase transitions were found at temperatures $T_{\mathrm{N1}}=6~\rm K$ and $T_{\mathrm{N2}}=5~\rm K$, respectively. Only part of the moments are statically ordered in this intermediate state between $T_{\mathrm{N1}}$ and $T_{\mathrm{N2}}$. Powder neutron diffraction experiments at different temperatures were performed as well. An incommensurate magnetic propagation vector of $\mathbf{k_{\rm m}}=(0.5,0.4,0.5)$ was identified. The refined spin configurations through the irreducible representation analysis confirmed that these Ising spins are canted in the crystal $ab$~plane.\end{abstract}

\pacs{75.47.Lx, 75.50.Ee, 75.40.-S, 75.40.Cx, 75.40.Gb,75.30.-m,75.30.Cr, 75.30.Gw}
\maketitle

\section{INTRODUCTION}
\maketitle
One-dimensional quantum magnets have received extensive research interest because of the emerging collective behaviors and strong quantum fluctuations enhanced by the low dimensionality~\cite{Intro_1,Intro_2}. Unique magnetic ground states and exotic spin dynamics, such as the fractional spinon excitations have been explored with thermal property studies and neutron scattering techniques~\cite{One_D_fraction_excitations_1,One_D_fraction_excitations_2,One_D_fraction_excitations_3,One_D_fraction_excitations_4,One_D_fraction_excitations_5,One_D_fraction_excitations_5}. In addition, an external magnetic field can be applied along the longitudinal and transverse directions, which plays another important role in exploring the new exciting physics in one-dimensional quantum magnets~\cite{Longi_field,Transverse_field}. Besides the experimental side of view, the most important part is that the one dimensional quantum spin systems provided a particular platform for testing various theoretical and numerical models~\cite{Haldane_gap,E8}. The possibility to compare the experimental results with the exact theoretical and numerical solutions allows very precise analysis of the exotic physics in one-dimensional systems, which makes it one of the most exciting topics in condensed matter physics in the last few decades~\cite{One_D_prog_1,One_D_prog_2,One_D_prog_3,One_D_prog_4,One_D_prog_4}.

The recent work on the one-dimensional antiferromagnetic Ising spin chain compounds BaCo$_2$V$_2$O$_8$~\cite{BaCoVO_1} and SrCo$_2$V$_2$O$_8$~\cite{SrCoVO_1,SrCoVO_2,SrCoVO_3,SrCoVO_4} presents a nice example. Exotic spin excitations involve two and more consecutive magnetic moments, which were known as Bethe strings~\cite{Bethe_strings}, have been experimentally observed and quantitatively characterized in SrCo$_2$V$_2$O$_8$~\cite{SrCoVO_5}. However, due to the large exchange interactions between the magnetic Co$^{2+}$ ions, high magnetic field larger than 20 T is required for further investigation of these exotic quantum excitations near the quantum critical region. Few experimental techniques have access to such a large magnetic field. On the contrary, the critical fields in one-dimensional quantum magnets with rare-earth ions are much smaller due to the relatively weak exchange interactions and the large effective $g$-factors arising from the strong spin orbital coupling. Relatively little research has been done on these systems, so new $f$-electron based one-dimensional quantum spin chains with accessible upper critical fields are highly demanded.

The rare-earth based germanide system RT$_5$Ge$_3$ (R = Ce, Nd, Dy, Yb, and T = Ni, Pd, Pt) crystallizes in an orthorhombic structure, with well separated one-dimensional spin chains, composed of the rare earth ions. For these rare-earth systems, due to the large splitting of the crystalline electrical field (CEF), the low temperature magnetism is usually dominated by the low lying CEF states. For Kramers ions, such as Ce, Nd, Dy and Yb, the CEF splitting normally leads to a doublet ground state described by pseudo-spin $S=1/2$ with anisotropic effective $g$-factors. The ground doublet state wave functions are determined by the surrounding charges and the local point symmetry. It is interesting to notice that, the crystal space group of RT$_5$Ge$_3$ and the local site symmetry of the rare-earth ions were identical to the rare earth based orthorhombic perovskite system RTO$_3$ (R=Ce-Lu, T=Fe, Al, Sc), where one dimensional quantum magnetism has been explored~\cite{DyScO3_PRB,YbAlO3_NC,YbAlO3_PRB}. Similar CEF configurations and one-dimensional magnetism are expected in this germanide system as well.

Previous studies on the isostructure compounds CePd$_5$Ge$_3$ and CePt$_5$Ge$_3$ reveal an antiferromagnetic ground state with significant spin fluctuations arising from the one dimensionality~\cite{CeP5Ge3_1,CeP5Ge3_2,CeP5Ge3_3,CeP5Ge3_4}. On the other hand, ferromagnetic like behaviors were observed for NdPd$_5$Ge$_3$ and NdPt$_5$Ge$_3$~\cite{NdP5Ge3_1}. Here, in this paper, we choose the Dysprosium (Dy) based germanide compound~\DNG, whose magnetic properties have not been explored yet. Single crystalline~\DNG~samples have been grown by the flux method, and thermal properties under field have been measured. An antiferromagnetic ground state is found. The zero field magnetic structure has been refined with the powder neutron diffraction measurements, and an incommensurate magnetic ground state is identified.

\section{EXPERIMENTAL DETAILS}
%Both single crystalline and polycrystalline samples of~\DNG~were prepared.
The single crystalline samples, used for thermal property measurements, were grown by the metallic flux method. High purity starting elements of Dysprosium (Dy), Nickel (Ni) and Germanium (Ge) were placed in alumina crucibles with excessive Bismuth (Bi) as flux. The assembly was sealed in evacuated quartz tubes, and heated up to 1050 $\rm^{\circ}C$ until all the starting materials were completely melted. The assembly was then slowly cooled down to 650 $\rm^{\circ}C$ at a rate of 3$\rm^{\circ}C/h$, at which temperature the extra Bi flux was mechanically removed by the centrifuge. Needle shaped~\DNG~single crystals of sizes about $0.1\times0.2\times5 \rm mm^3$ were left inside the growth crucibles. The crystal structure of~\DNG~was verified by the Bruker APEX-II CCD diffractometer. The magnetic susceptibility and specific heat measurements were performed by the commercial Quantum Design Magnetic Property Measurement System (MPMS) and the Physical Property Measurement System (PPMS).

To identify the zero field magnetic structure, an additional 3 g polycrystalline sample of~\DNG~was synthesized by arc-melting a stoichiometric amount of the high purity starting elements under argon atmosphere. Powder neutron diffraction measurements were performed by the time-of-flight powder diffractometer, POWGEN, at Spallation Neutron Source (SNS), Oak Ridge National Laboratory~\cite{Powgen}. An orange cryostat was used as the sample environment to cover a temperature region down to 1.8~K. High resolution neutron diffraction patterns were collected by two neutron frames with center wavelengths of 1.5 and 2.665~$\rm\AA$.

\section{Results and Analysis}

\subsection{Crystal Structure and CEF}

\begin{figure}[ht!]
 \includegraphics[width=3in]{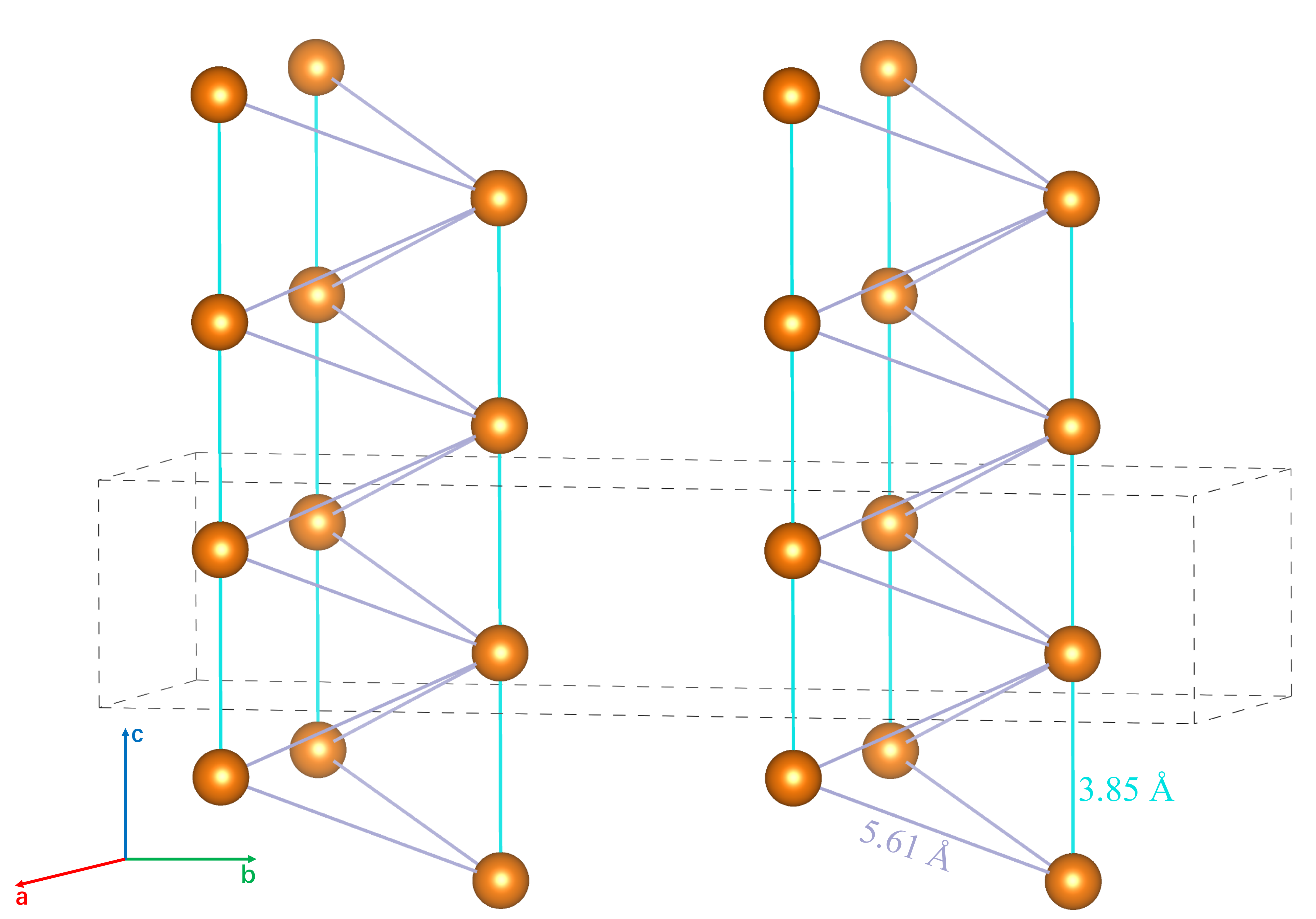}
    \caption{Simplified crystal structure of \DNG, where only Dy ions were shown. With the Pbnm notation, the Dy ions form one-dimensional chains along the crystal $c$~axis.}
\label{structure}
\end{figure}

\begin{table}[h]
\caption{Refined parameters and crystallographic data of~\DNG~from single crystal X-ray diffraction pattern.}
\centering
\begin{tabular}{p{5.2cm}<{\raggedright}p{3cm}<{\raggedright}}
\hline
~~~~Chemical Formula                 & \DNG\\
~~~~Formula Weight                   & 673.82\\
~~~~Temperature(K)                   & 100\\
~~~~$\lambda(\rm\AA)$                   & 0.71073\\
~~~~Crystal system                   & Orthogonal\\
~~~~Space group, $\textit{Z}$        & Pbnm(No.62), 4\\
~~~~$a(\rm\AA)$                           & 6.7934(3)\\
~~~~$b(\rm\AA)$                           & 19.0737(11)\\
~~~~$c(\rm\AA)$                           & 3.8549(2)\\
~~~~$\alpha,\beta,\gamma(\rm^{\circ})$    & 90\\
~~~~$V(\rm\AA^{3})$                       & 499.50(4)\\
~~~~$\rho_{\rm calc}(\rm g/cm^{3})$           & 8.96\\
~~~~Goodness-of-fit                    & 1.109\\
~~~~$R_{1}$                            & 0.0484\\
~~~~$wR_{2}$                           & 0.1326\\
\hline % inserts single-line
\end{tabular}

\label{tab:structure1}
\end{table}

\begin{table}[h]
\caption{Atomic displacement parameters of \DNG.}%title of the table
\centering % centering table
\begin{tabular}{p{0.8cm}<{\centering}p{0.6cm}<{\centering}p{1.9cm}<{\centering}p{1.9cm}<{\centering}p{1cm}<{\centering}p{1.6cm}<{\centering}}
\hline
Atom & Wkf. & x  & y & z & U$_{eq}$\\
\hline
Dy    & 4c  & 0.878101 & 0.355457 & 0.25  &0.021540\\
Ge1   & 4c  & 0.648335 & 0.075853 & 0.25  &0.022257\\
Ge2   & 4c  & 0.091645 & 0.086486 & 0.25  &0.022173\\
Ge3   & 4c  & 0.382245 & 0.262429 & 0.25  &0.021753\\
Ni1   & 4c  & 0.365927 & 0.003724 & 0.25  &0.021713\\
Ni2   & 4c  & 0.072046 & 0.207525 & 0.25  &0.022396\\
Ni3   & 4c  & 0.689564 & 0.199837 & 0.25  &0.022187\\
Ni4   & 4c  & 0.358735 & 0.386945 & 0.25  &0.023130\\
Ni5   & 4c  & 0.616633 & 0.486552 & 0.25  &0.022110\\
\hline % inserts single-line
\end{tabular}

\label{tab:structure2}
\end{table}

\DNG~crystallized in an orthorhombic YNi$_5$Si$_3$-type structure with the space group No.62. The crystal structure has been verified by the single crystal X-ray diffraction measurement. The lattice parameters refined at 100 K are found to be $\alpha=\beta=\gamma={90}{^{\circ}}$, and $a={6.7934(3)}{~\rm\AA}$, $b={19.0737(11)}{~\rm\AA}$, $c={3.8549(2)}{~\rm\AA}$, with the conventional $P\rm bnm$ notation (Table \ref{tab:structure1} and \ref{tab:structure2}). We choose this traditional $P\rm bnm$ notation for reasons. An advantage is that the one-dimensional Dysprosium (Dy$^{3+}$) spin chain is able to be along the crystal $c$~axis. The distance between the nearest neighbour ions along the chain direction is about 3.85~\AA, while the inter-chain distance is about 5.61~\AA, as indicated in Fig.~\ref{structure}. In addition, the crystal $ab$~plane is exactly the mirror plane with this $P\rm bnm$ notation. This will make the discussion about the crystal field more straightforward, and make these results obtained here comparable with other systems sharing the same local point symmetry.

In \DNG, the magnetic Dysprosium Dy$^{3+}$ atoms are surrounded by the nearby Ni and Ge charges, with a local point symmetry $C_{\rm s}(m)$. This low local point symmetry would then lift the $2J+1=16$ ($L=5, S=5/2$, and $J=15/2$) multiplet states of Dy$^{3+}$ into eight doublet CEF states because of the strong spin-orbital couplings. The mirror symmetry in the crystal $ab$~plane usually results in an Ising like ground state, with easy moment direction constrained either along the crystal $c$~axis, which is perpendicular to the mirror plane, or lying within the $ab$ mirror plane. In the latter case, the magnetic moments are likely to lie along the direction of an angle $\varphi$ with respect to the $a$ or $b$~axis of the principle crystal, rather than strictly along them. This is because no additional rotational symmetry is preserved in this $ab$ mirror plane. In addition, the local charge environments of the Dy$^{3+}$ ions from nearby chains are connected by an inversion symmetry operation. This inversion symmetry operation relates the Ising moments of nearby chains with canted angles $\pm\varphi$ from the crystal $a$ or $b$~axis symmetrically. This situation is identical to that in the rare-earth orthorhombic perovskite oxides such as DyScO$_3$ and YbAlO$_3$~\cite{DyScO3_PRB,YbAlO3_NC,YbAlO3_PRB}. However, different from those perovsike oxides, the compound~\DNG\ is metallic, and it is hard to assign an accurate charge to the surrounding Ni and Ge ions, due to the screen effect of the conducting electrons. Thus, the CEF calculations based on the point charge model can only give qualitative guide, rather than an accurate prediction as in rare earth oxides~\cite{DyScO3_PRB,YbAlO3_NC,YbAlO3_PRB}. Therefore, to examine the magnetic ground state, further experimental characterizations with the anisotropic magnetization and neutron diffraction experiments are required.
%As the measuring temperatures were lower than the first exited CEF energy levels, the magnetic properties would then be dominated by the ground doublet states.

\subsection{Magnetization and Single Ion Anisotropy}

\begin{figure}[ht!]
 \includegraphics[width=3in]{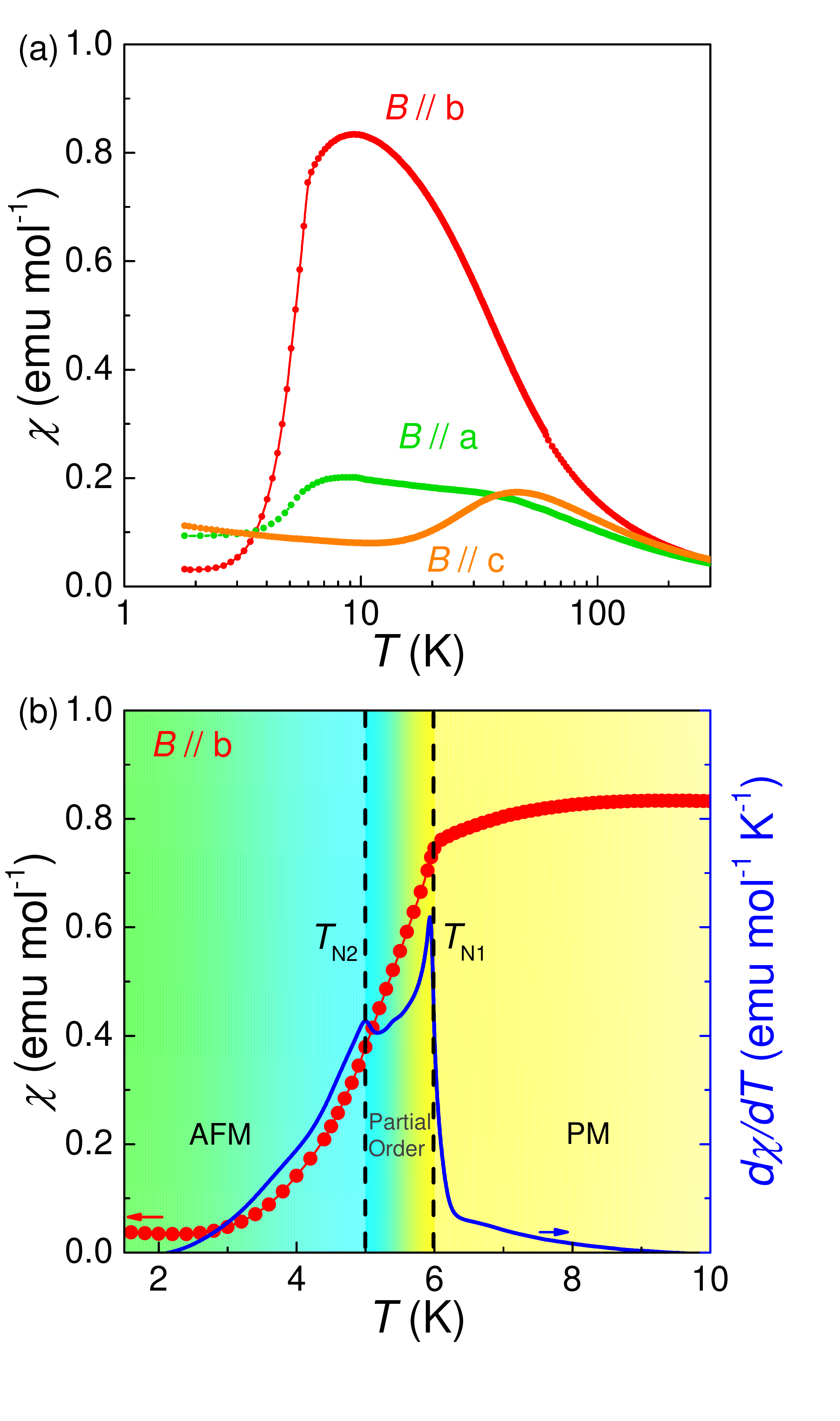}
    \caption{(a) Temperature-dependent DC magnetic susceptibility $\chi$ ( measured at $B=0.1$ T) with external magnetic fields applied along different principle axis.  (b) The temperature-dependent magnetic susceptibility and the temperature derivative of the susceptibility ($d\chi\rm/dT$) at temperatures lower than 8 K, measured with field applied along the $b$~axis. Two antiferromagnetic transitions at temperatures $T_{\mathrm{N1}}=6 \rm K$ and $T_{\mathrm{N2}}=5 \rm K$ were observed, as indicated by the vertical dashed lines. }
\label{MT}
\end{figure}

\begin{figure}[ht!]
 \includegraphics[width=3.2in]{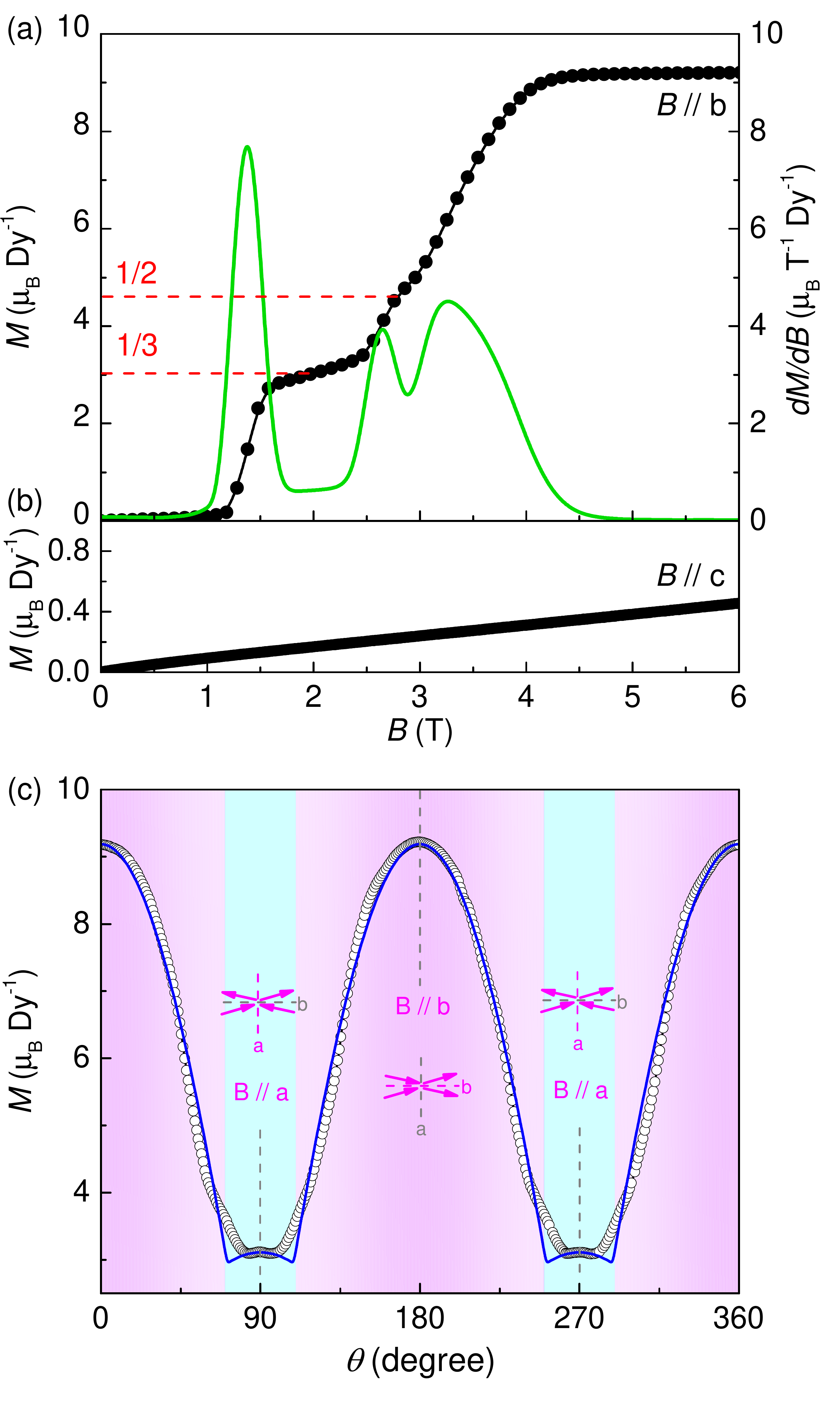}
    \caption{(a) The field-dependent magnetization, measured at T = 1.8 K, with the field applied along $b$ (within the mirror plane), (b) and $c$~axis (perpendicular to the mirror plane). (c) The angle-dependent magnetization measured at $T = 5 $ K with the magnetic field applied in the $ab$ mirror plane. The blue line represents the fitting assuming two canted Ising magnetic lattices.}
\label{MH}
\end{figure}

The temperature-dependent magnetic susceptibility $\chi$ with field along different directions is shown in Fig.~\ref{MT}a. We can see that these measured susceptibility begin to deviate from each other below about 200 K. This single ion anisotropy is probably due to the CEF effect. When the measuring temperatures are lower than the excited CEF states, the magnetic properties will be dominated by the ground doublets. Depending on the wave functions of the ground doublets, strong single ion anisotropic behaviors start to build up. In Fig.~\ref{MT}a, we found that the measured magnetic susceptibility in the $ab$~plane becomes larger than that along $c$~axis as the temperature decreases. This suggests that these Dy moments are lying in the $ab$-mirror plane, as in systems like DyScO$_3$~\cite{DyScO3_PRB}. In addition, the magnetization along $b$~axis seems to be the dominant one, indicating that the local easy axis is close to the crystal $b$~axis. Additional broad peaks are observed around 40 K for $B\parallel a$ and $B\parallel c$. These are probably due to the contributions of the excited CEF levels, which may locate at the energy scale of about 40 K to 200 K (or roughly $4\sim 20$ meV).

It is the magnetic properties at the lowest temperatures that matter here, which are dominated by the ground doublet states. Thus, to avoid the complexity, we focus on the magnetization within the temperature region below about 10 K. As shown in Fig.~\ref{MT}b, the low temperature magnetization for $B\parallel b$ is measured in 0.1 T. A broad maximum like feature around 9 K was observed in $\chi(T)$, indicating short range correlations already start to build up. Following this broad maximum like feature, a significant drop at $T_{\rm N1}=6$ K is observed, indicating a long range antiferromagnetic order. In addition, in the temperature derivative of the susceptibility ($d\chi/dT$) (blue line in Fig.~\ref{MT}b), an additional peak like anomaly shows up at $T_{\rm N2}=5$ K. Similar phenomena have been observed in other low dimensional frustrated metallic systems before~\cite{HoAgGe}. In the case here, if next near neighbor interactions between chains are included, these one dimensional spin chains actually form two dimensional anisotropic triangular lattices (gray lines in Fig.~\ref{structure}). In this anisotropic triangular lattice, Dy spin chains are connected into bulked layers, composed of isosceles triangles. Considering that~\DNG~is a metallic system with large Ising moments, neither the long range Ruderman-Kittel-Kasuya-Yosida (RKKY) interaction nor the dipolar interaction can be neglected. Thus, the spin frustration is induced by these long range inter-chain interactions. In the mean time, part of the frustration is released with the establishment of a long range statical order. As proposed in HoAgGe~\cite{HoAgGe}, a possible scenario is that the antiferromagnetic ordering at $T_{\rm N1}=6$ K is not a complete phase transition, and the system is in a partially ordered state in between $T_{\rm N1}$ and $T_{\rm N2}$. This scenario will be further discussed together with the specific heat and neutron scattering results.

%We have to emphasize here that the notation Pbnm was used for all the data in this report.

The field-dependent magnetization is measured as presented in Fig.~\ref{MH}. Significant anisotropic behaviors are observed when an external field is applied in the $ab$ plane and along the crystal $c$ axis. Plateau like anomalies at $1/3$ and $1/2$ are observed at 1.8 K as $B\parallel b$. These fractional plateau phases are also evident by the dip like features in the field derivative magnetic susceptibility ($dM/dB$) (green line in Fig.~\ref{MH}a). As discussed previously, these plateau like states are induced by the spin frustration. These observations further confirm that multiple inter-chain interactions are playing important roles in this system. On the other hand, linear like field-dependent magnetization is observed as $B\parallel c$ (Fig.~\ref{MH}b). The high-field saturation moment is about $9.2~\mu_{\rm B}\rm /Dy$ for $B\parallel b$, which is one order larger than the measured magnetic moment along the $c$ axis. These behaviors confirm that these Dy$^{3+}$ magnetic moments are lying in the $ab$-mirror plane.

As expected from the CEF analysis based on the point symmetry, these moments are likely to be canted from the crystal $a$ or $b$ axis. To verify this, we performed the measurements of the angle-dependent magnetization with field rotating in the $ab$~plane and the result is shown in Fig.~\ref{MH}c. These angles at $90^{\rm o} (270^{\rm o})$ and $0^{\rm o} (180^{\rm o})$ correspond to $B\parallel a$ and $B\parallel b$, respectively. We find that dip-like features appear symmetrically around  $90^{\rm o}$ and $270^{\rm o}$. With the assumption of two canted magnetic sublattice of Dy$^{3+}$ Ising moments, the measured angle dependence can be described as:
\begin{align}
\label{MvsB}
M&=\dfrac{M_{\rm s}}{2}(\lvert \rm cos(\varphi-\theta)\rvert+\lvert \rm cos(\varphi+\theta)\rvert).
\end{align}
Here $M_{\rm s}$ is the saturation moment, $\varphi$ is the local easy direction of these Ising Dy$^{3+}$ moments from the $b$~axis, and $\theta$ is the angle between the external field and crystal $b$ axis. The fitting is represented as the blue line shown in~Fig.~\ref{MH}c. The extracted saturation moment is $M_{\rm s}=9.7~\mu_{\rm B}\rm /Dy$, with titling angle $\varphi=18.7^{\rm o}$. These values are well consistent with the field-dependent magnetization shown in Fig.~\ref{MH}a, where the saturation moments for $B\parallel b$ equal to $M_{\rm s} (B\parallel b)=9.7\times\rm cos\varphi\simeq9.2~\mu_{\rm B}\rm /Dy$.

\subsection{Specific Heat and Integrated Entropy}
\begin{figure}[ht!]
 \includegraphics[width=2.65in]{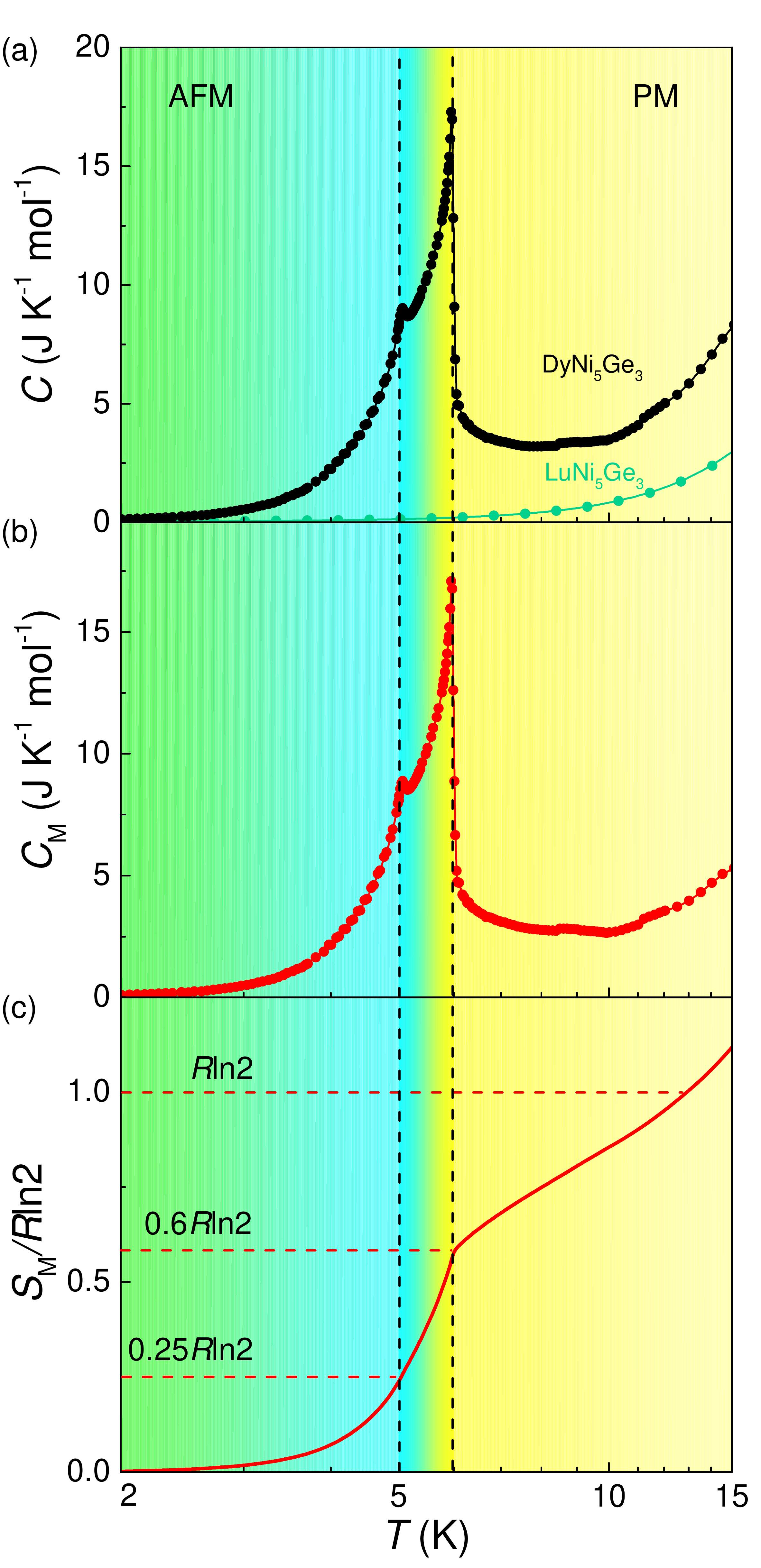}
    \caption{(a) The temperature-dependent specific heat $C$ of ~\DNG (black line) and  LuNi$_5$Ge$_3$ (green line) in zero fields. (b) The corrected magnetic specific heat and (c)the integrated temperature-dependent entropy. Vertical dashed lines indicate the two successive phase transitions at $T_{\rm N1}=6$ K and $T_{\rm N2}=5$ K. }
\label{HC}
\end{figure}

The temperature-dependent specific heat $C$ is measured in zero field and presented in Fig.~\ref{HC}a. The nonmagnetic isostructural compound LuNi$_5$Ge$_3$ is synthesised as well, and the specific heat of LuNi$_5$Ge$_3$ is used as the estimation of the lattice contributions (green circles in Fig.~\ref{HC}a). The corrected magnetic specific heat $C_{\rm M}$ is then shown in Fig.~\ref{HC}b. Two subsequent phase transitions were observed at $T_{\rm N1}=6$ K and $T_{\rm N2}=5$ K in the temperature-dependent specific heat. These transition temperatures are well consistent with the observations in the magnetization derivatives $dM/dT$, as presented in Fig.~\ref{MT}b.

The integrated magnetic entropy is presented in Fig.~\ref{HC}c. A full entropy of $R$ln2 is realized around 10 K, confirming a doublet ground state at lower temperatures. Generally speaking, for classical magnets, the critical fluctuations are weak, and the full entropy or at least 80-90\% of the full entropy will be released at the transition temperature~\cite{Yb3Pt4}. Here, in contrast, we found that only about 25\% of the full entropy is released at $T_{\rm N2}=5$ K and about 60\% of the full entropy is released at $T_{\rm N1}=6$ K for~\DNG. There are still about 40\% of the full entropy left above the ordered state, indicating the existence of strong spin fluctuations. It is also noticed that the entropy of $60\%R$ln2 related to the ordered state is released successfully through the two transitions.These observations suggest that the magnetic Dy$^{3+}$ moments are not completely ordered at $T_{\rm N1}$. The coexistence of a great amount of fluctuations suggest that these moments are likely only partially frozen in the intermediate ordered state between $T_{\rm N1}$ and $T_{\rm N2}$. What's more, the additional magnetic contributions larger than $R$ln2 is observed at temperatures higher than 10 K, even with the lattice contribution removed. The magnetic entropy may be contributed from the excited CEF levels, which may locate at energy levels around a few milli-electronvolts.

\subsection{Neutron Scattering and Magnetic Ordering }
% plot: neutrons.
% order parameter, contour plot of the temperature dependence,
% refined magnetic structure (help from Jieming)
\begin{figure}[ht!]
\flushleft
 \includegraphics[width=3.5in]{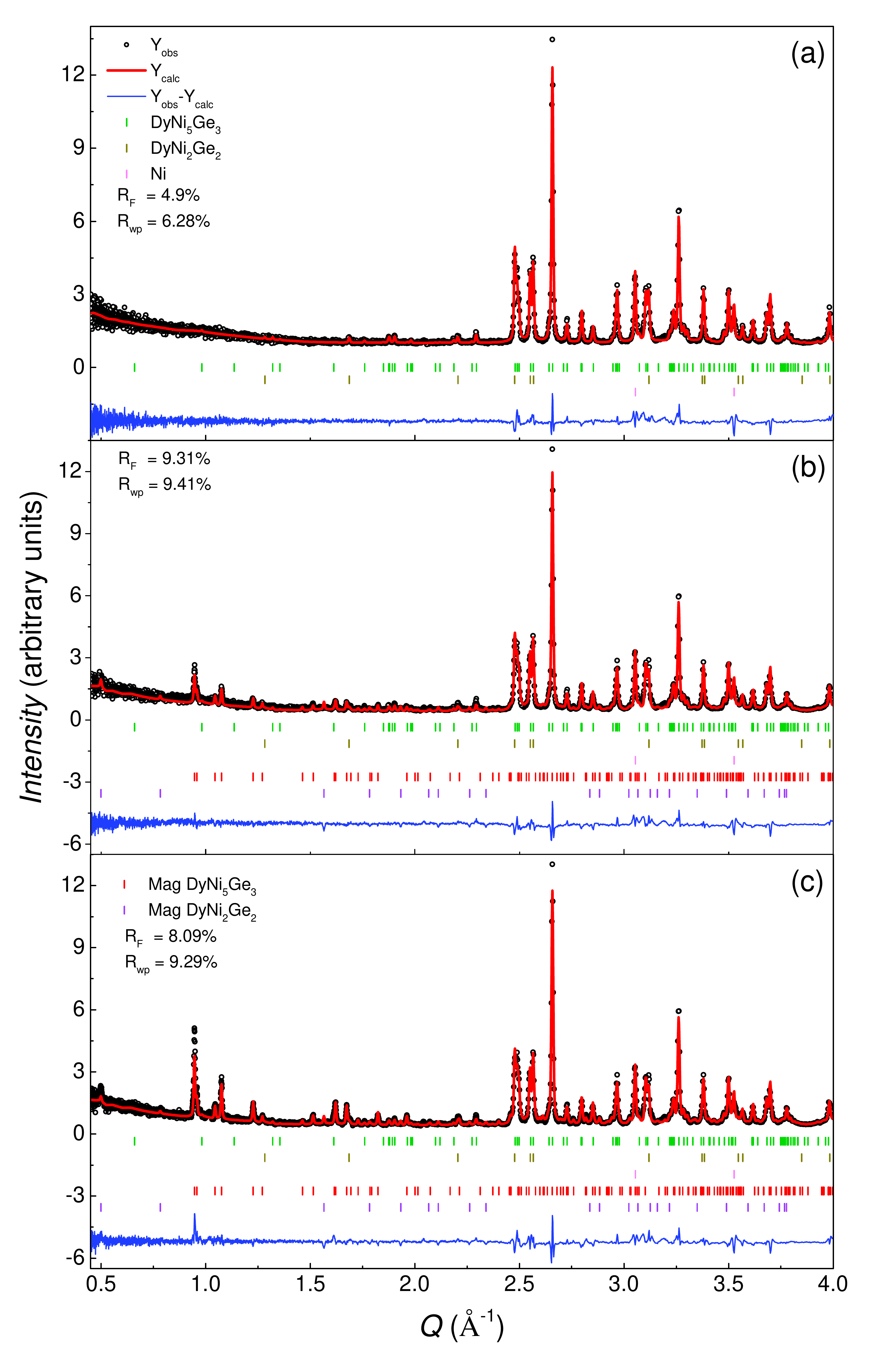}
    \caption{Rietveld refinement of neutron powder diffraction patterns
    of \DNG at (a) 10 K, (b) 5.5 K, and (c) 1.8 K.
    The black points represent actual data and red lines represent
    the Rietveld fitting to the data. The difference curve is shown at the bottom.}
\label{mag_diffraction}
\end{figure}

\begin{figure}[ht!]
\flushleft
 \includegraphics[width=3.4in]{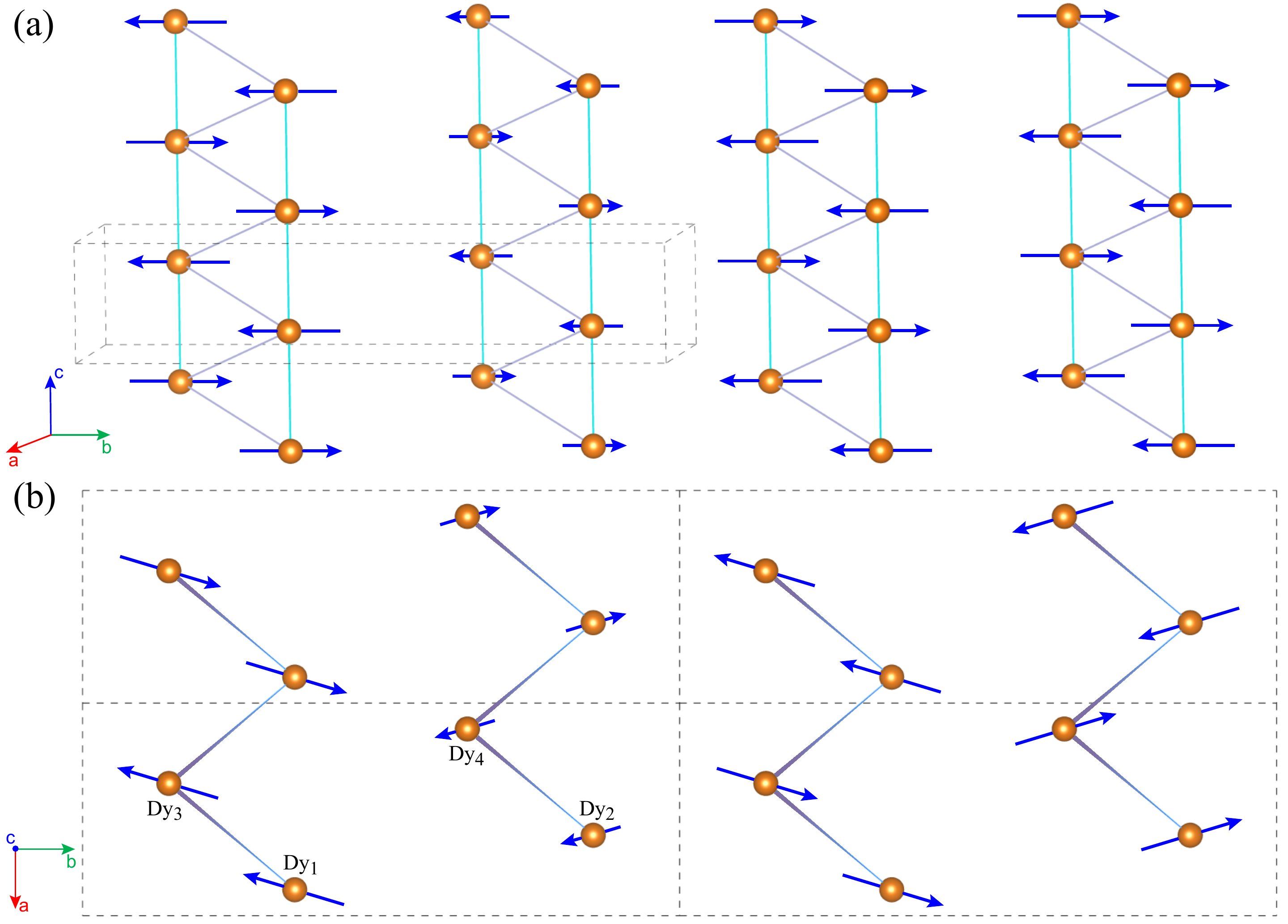}
    \caption{The magnetic structure of~\DNG~below the ordering temperature
    refined from the neutron data. The arrows indicate the ordered Dy$^{3+}$ spins,
    which are found to lie in the ab~plane.}
\label{mag_structure}
\end{figure}

\begin{figure}[b]
 \includegraphics[width=2.8in]{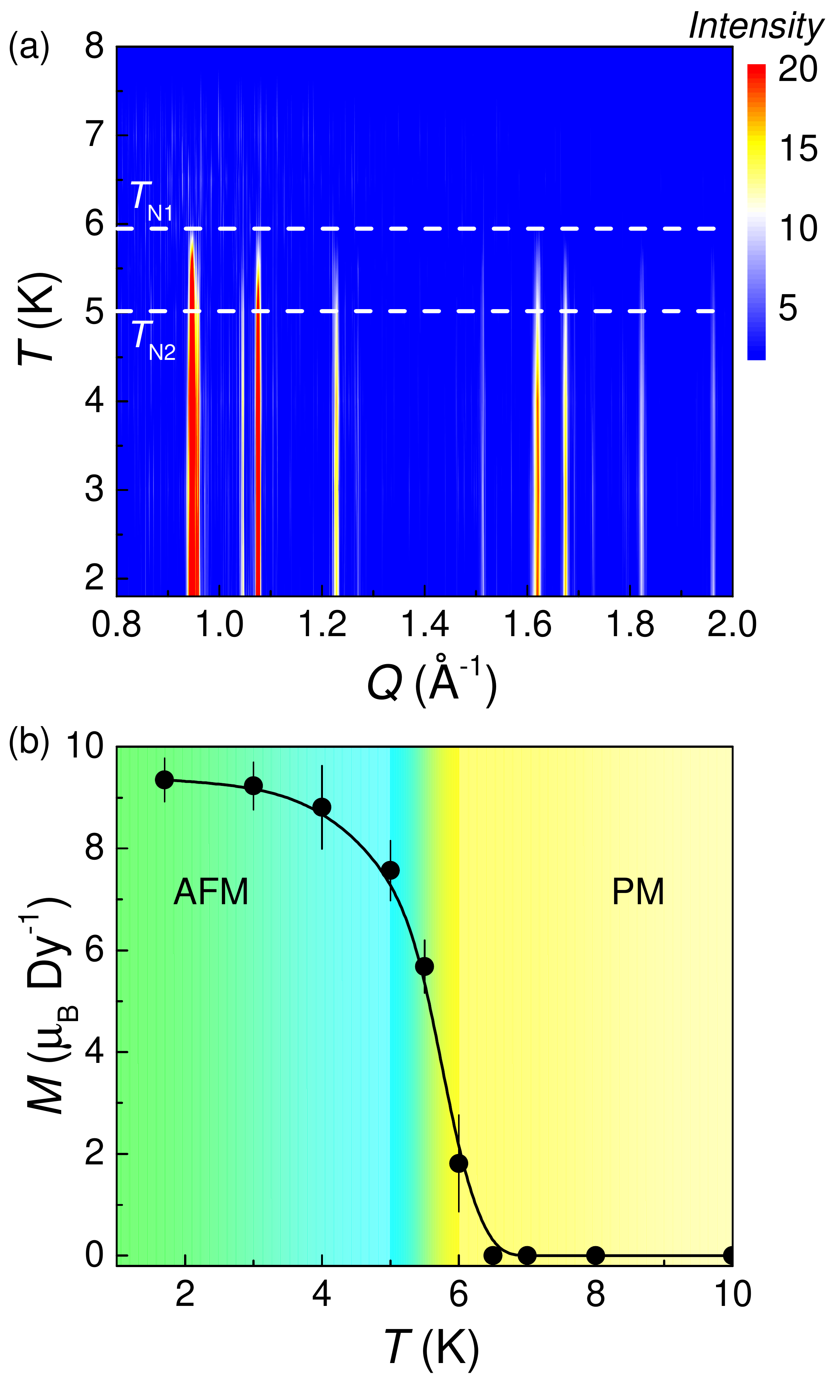}
    \caption{(a) The contour plot of the measured neutron diffraction pattern collected at POWGEN in frame 3 as functions of temperature $T$ and wave vector $Q$ of the~\DNG~powdered sample. (b) Temperature evolution of the intensity of refined moment.}
\label{order_parameter2}
\end{figure}

To determine the magnetic structure of DyNi$_5$Ge$_3$, powder neutron diffraction measurements have been performed with the time-of-flight powder diffractometer POWGEN~\cite{Powgen}. Fig.~\ref{mag_diffraction}a-c present the powder diffraction patterns measured at 10~K, 5.5~K and 1.8~K. Consistent with our single crystal X-ray diffraction results, the Rietveld refinement of the neutron diffraction data at 10 K confirms the orthorhombic structure of DyNi$_5$Ge$_3$ with the space group $P\rm bnm$ (No.62). These powder samples for neutron scattering were prepared separately by the arc-melt method. Besides the main phase of DyNi$_5$Ge$_3$, about $15.57\pm0.18\%$ DyNi$_2$Ge$_2$ and $7.12\pm0.22\%$ Ni are identified as additional impurity phases inside the powder sample. For $T=1.8$ K, a new set of reflections emerged around the low-$Q$ region, indicating the magnetic origin. By applying the $K$-search program of the FullProf Suite, an incommensurate magnetic propagation vector, $\mathbf{k_{\rm m}}$=(0.5,0.4,0.5), can be determined for DyNi$_5$Ge$_3$. We also found that there was an extra weak magnetic peak around $Q=0.499$ \AA$^{-1}$, which could be indexed by the magnetic propagation vector $\mathbf{q_{\rm m}}$=(0,0,0.779) for the impurity DyNi$_2$Ge$_2$ phase. This propagation vector is consistent with the previously reported powder neutron diffraction results of DyNi$_2$Ge$_2$~\cite{DyNi2Ge2_1,DyNi2Ge2_2}.

\begin{table*}[ht!]
\renewcommand\arraystretch{1.5}
\begin{ruledtabular}
\caption{Basis vectors of the irreducible representations (IR) for DyNi$_5$Ge$_3$ with magnetic wave-vector $\mathbf{k_{\rm m}} =(0.5,0.4,0.5)$ and Dy site at (0.877,0.355,0.250).}
\centering
\begin{tabular}{lcccccc}
   IR           & Basis vector   &  Dy$_1$ (0.877,0.355,0.250) & Dy$_2$ (0.623,0.855,0.250) & Dy$_3$ (0.377,0.145,0.750) & Dy$_4$ (0.123,0.645,0.750) \\
                &                &   $m_{x}$   $m_y$    $m_z$   &   $m_x$   $m_y$    $m_z$  &   $m_x$   $m_y$    $m_z$  &  $m_x$   $m_y$    $m_z$   \\
\hline
    $\Gamma_1$  &  $\psi_1$      &      0  0  1               &   0   0  -0.309+0.951i    &        0  0  1            &    0   0  -0.309+0.951i   \\
\hline
     $\Gamma_2$ &  $\psi_2$      &      1  0  0               &   -0.309+0.951i  0  0     &        1  0  0            &   -0.309+0.951i  0  0     \\
                &  $\psi_3$      &      0  1  0               &   0   0.309-0.951i  0     &        0  1  0            &   0   0.309-0.951i  0     \\
\hline
     $\Gamma_3$ &  $\psi_4$      &      0  0  1               &   0   0   0.309-0.951i    &        0  0  1            &   0   0   0.309-0.951i    \\
\hline
     $\Gamma_4$ &  $\psi_2$      &      1  0  0               &   0.309-0.951i   0  0     &        1  0  0            &   0.309-0.951i   0  0     \\
                &  $\psi_3$      &      0  1  0               &   0   -0.309+0.951i  0    &        0  1  0            &   0  -0.309+0.951i  0     \\
\end{tabular}
\end{ruledtabular}

\label{tab_BV}
\end{table*}

Representation analysis is applied to find the possible magnetic structures for DyNi$_5$Ge$_3$~\cite{Diffraction_mag_stru_determ}. For Dy$^{3+}$ site at (0.877,0.355,0.250) and the magnetic propagation vector $\mathbf{k_m}$=(0.5,0.4,0.5), the possible spin configurations can be described by four different irreducible representations (IRs). Table \ref{tab_BV} lists the basis vectors for IRs $\Gamma_1$, $\Gamma_2$, $\Gamma_3$ and $\Gamma_4$, respectively. For $\Gamma_1$ and $\Gamma_3$, the moments are constrained along the $c$ axis, which are in contrast to the magnetization measurement results. For $\Gamma_2$ and $\Gamma_4$, the moments are allowed to lie in the $ab$ plane. By Rietveld refinement, we found that $\Gamma_4$ could give the best fit with $R_{\rm F} = 8.09 \%$ and $R_{\rm wp} = 9.29 \%$ for the powder neutron diffraction data at 1.8 K, as shown in Fig.~\ref{mag_diffraction}c. The determined magnetic structure is presented in the Fig.~\ref{mag_structure}a and b. All the moments are lying in the $ab$ plane. The spins on site Dy$_1$ and Dy$_3$, Dy$_2$ and Dy$_4$ are aligned in parallel, with a tilting angle $\pm 13.95(5)^{\rm o} $ from the $b$~axis. The tilting angle $13.95^{\rm o}$ refined from neutron diffraction experiments is slightly different from the angle of $18.7^{\rm o}$ determined through the previous angle-dependent magnetization measurement. However, considering that these are two independent measurements with their own systematic errors, we think these two values are still in good agreement with each other. Based on the propagation vector $\mathbf{k_m}$=(0.5,0.4,0.5), each spin of Dy$^{3+}$ in the crystal unit cell is antiferromagnetically aligned along the $a$ and $c$~axis, and the amplitude of the spins is modulated by $\mathbf{k_y} \cdot \mathbf{r}$ along the $b$~axis. The amplitude moment for each Dy ion is about 9.41(39)~$\mu_{\rm B}$.

Fig.~\ref{order_parameter2}a shows the contour plot of the measured neutron diffraction pattern as functions of temperatures. New set of magnetic reflections begins to develop below the antiferromagnetic ordering temperature $T_{\rm N1}=6$ K. However, no additional set of reflections is found below the second transition temperature at $T_{\rm N2}=5$ K. With further decreasing temperature, the intensity of all the magnetic peaks saturates below $T_{\rm N2}$. The temperature dependence of the refined moment of Dy$^{3+}$ is shown in Fig.~\ref{order_parameter2}b, which is consistent with the evolution of the magnetic peaks. These neutron scattering results suggest that the Dy$^{3+}$ moments in~\DNG~may share the same ordering pattern in the intermediate temperature region between $T_{\rm N2}$ and $T_{\rm N1}$ and the lower temperature phase below $T_{\rm N2}$~(Fig.~\ref{mag_diffraction}b and c). It is possible that the moments are only partially ordered in the intermediate temperature between $T_{\rm N2}$ and $T_{\rm N1}$, and then got further frozen in the phase below $T_{\rm N2}$. This is also consistent with the observations found in the specific heat and entropy measurements.

\section{Conclusion}
In conclusion, we have performed a detailed investigation of~\DNG~through the specific heat, magnetization and powder neutron diffraction measurements. The crystal electrical field based on the crystal structure and local point symmetry suggests an Ising-like ground double state. Additional angle-dependent magnetization in the crystal $ab$~plane reveals that these Ising Dy$^{3+}$ moments are canted from the $b$~axis by angle $\varphi\simeq18.7^{\rm o}$. The temperature-dependent magnetization and specific heat reveal two successive antiferromagnetic transitions at $T_{\rm N1}=6$ K and $T_{\rm N2}=5$ K. In the ordered state, fractional plateau phases at $M/M_{\rm s}=1/3$ and  $M/M_{\rm s}=1/2$ are observed with magnetic field applied along the $b$ axis. The measured specific heat and integrated magnetic entropy suggest that there exist a great amount of fluctuations persistent in the intermediate phase between $T_{\rm N1}$ and $T_{\rm N2}$. In addition, magnetic reflections are observed below $T_{\rm N1}=6$ K in powder neutron scattering experiments. It is interesting to find that no additional magnetic peaks appeared below $T_{\rm N2}=5$ K. A possible scenario is that the intermediate phase and the lower temperature magnetic phase below $T_{\rm N2}=5$ K share a similar magnetic structure , but only partially ordered moments is observed in the former. The ground state magnetic structure has been presented as well with the irreducible representations analysis. The refined magnetic moments are found out to be tilted from the $b$ axis by about $\varphi\simeq13.95^{\rm o}$. This value is close to the tilting angle $\varphi\simeq18.7^{\rm o}$ extracted from the angle-dependent magnetization measurements. In the future, single crystal neutron diffraction measurements with fields are needed, and the magnetic configurations with these fields induced plateau phases are worth further exploring.

\begin{acknowledgments}
The research at SUSTech was supported by the National Natural Science Foundation of China (No.~12134020 ).
Part of this work was also supported by the National Natural Science Foundation of China (No.~11974157, No.~11875265 and No.~12104255.
This work was also supported by the Guangdong Basic and Applied Basic Research Foundation (No.~2021B1515120015), the Scientific Instrument Developing Project of the Chinese Academy of Sciences ($^{3}$He based neutron polarization devices), the Institute of High Energy Physics, the program for Guangdong Introducing Innovative and Entrepreneurial Teams (No.~2017ZT07C062), by Shenzhen Key Laboratory of Advanced Quantum Functional Materials and Devices (No.~ZDSYS20190902092905285).
The authors acknowledge the assistance of SUSTech Core Research Facilities.
Neutron diffraction measurements used resources at the Spallation Neutron Source, a DOE Office of Science User Facility operated by the Oak Ridge National Laboratory.
\end{acknowledgments}

\end{document}